\documentclass[reprint, 10pt, amsmath, amssymb, floatfix, aps, prl,  superscriptaddress, nofootinbib, nobibnotes,twocolumn]{revtex4-1}

\pdfoutput=1
\usepackage{graphicx}
\graphicspath{{figures/}}
\usepackage[utf8]{inputenc}
\usepackage{amsmath}
\usepackage{amsfonts}
\usepackage{amssymb}
\usepackage{multirow}
\usepackage{CJKutf8}
\usepackage{color}
\usepackage[colorlinks=true, linkcolor=red, citecolor=blue]{hyperref}
\usepackage[capitalise]{cleveref}
\usepackage{acro}
\usepackage{adjustbox}
\usepackage{array}
\usepackage{natbib}
\usepackage{dblfnote}
\DFNalwaysdouble
\usepackage{slashed}
\usepackage{dcolumn}
\usepackage{hyperref}
\usepackage{geometry}
\usepackage{inputenc}

\def\({\left(}
\def\){\right)}
\def\[{\left[}
\def\]{\right]}
\def\be{\begin{eqnarray}}
\def\ee{\end{eqnarray}}
\def\ss{\small}

\DeclareAcronym{GW}{
  short = GW ,
  long = gravitational wave ,
  short-plural = s 
}
\DeclareAcronym{LIGO}{
  short = LIGO ,
  long = Laser Interferometer Gravitational-wave Observatory ,
  short-plural = 
}
\DeclareAcronym{LISA}{
  short = LISA ,
  long = Laser Interferometer Space Antenna ,
  short-plural =  
}
\DeclareAcronym{SKA}{
  short = SKA ,
  long = Square Kilometre Array ,
  short-plural =  
}
\DeclareAcronym{FLRW}{
  short = FLRW ,
  long = Friedmann-Lemaitre-Robertson-Walker ,
  short-plural =  
}

\begin{document}

\title{Gauge Invariant Second Order Gravitational Waves}

\author{Zhe Chang}
\email{changz@ihep.ac.cn}
\affiliation{Theoretical Physics Division, Institute of High Energy Physics, Chinese Academy of Sciences, Beijing 100049, People's Republic of China}
\affiliation{School of Physical Sciences, University of Chinese Academy of Sciences, Beijing 100049, People's Republic of China}
\author{Sai Wang}
\email{wangsai@ihep.ac.cn}
\affiliation{Theoretical Physics Division, Institute of High Energy Physics, Chinese Academy of Sciences, Beijing 100049, People's Republic of China}
\affiliation{School of Physical Sciences, University of Chinese Academy of Sciences, Beijing 100049, People's Republic of China}
\author{Qing-Hua Zhu}
\email{zhuqh@ihep.ac.cn}
\affiliation{Theoretical Physics Division, Institute of High Energy Physics, Chinese Academy of Sciences, Beijing 100049, People's Republic of China}
\affiliation{School of Physical Sciences, University of Chinese Academy of Sciences, Beijing 100049, People's Republic of China}


\begin{abstract}
We investigate the gauge invariance of the second order gravitational waves induced by the first order scalar perturbations by following the Lie derivative method. It is shown explicitly that the second order gravitational waves are gauge invariant in the synchronous frame. In the gauge invariant framework, we derive the equation of motion of the second order gravitational waves and show that the second order gravitational waves are sourced from the first order scalar perturbations described well in the gauge invariant Newtonian frame. Since the observables of gravitational waves are measured in the synchronous frame, we define the energy density spectrum of the second order gravitational waves in terms of the gauge invariant synchronous variables. This way guarantees no fictitious tensor perturbations. It is shown that the gauge invariant energy density spectrum of the second order gravitational waves coincides with the one in the Newtonian gauge. 
\end{abstract}

\maketitle
\acresetall

\emph{Introduction.}---One of the most successful predictions of the inflation theory is the production of cosmological perturbations \cite{Mukhanov:1990me,Riotto:2002yw}. 
Due to precise measurements of the cosmic microwave background and the large scale structure, we have learnt a lot of information about the scalar and tensor perturbations on large scales \cite{Aghanim:2018eyx,Akrami:2018odb,Zyla:2020zbs}. 
However, we know little about them on smaller scales \cite{Emami:2017fiy,Gow:2020bzo}, due to the Silk damping of radiations and the nonlinear clustering of matters. 
In fact, the small-scale perturbations play important roles in understanding the inflationary physics and the dark matter. 
On the one hand, they could reflect a global evolution of the inflaton potential and the high energy physics behind it \cite{Braglia:2020eai,Byrnes:2018txb,Kawasaki:2019hvt,Ballesteros:2020sre,GarciaBellido:1996qt,Kohri:2007qn}, which are not completely accessed to for the large-scale ones. 
On the other hand, if enhanced, they can produce the primordial black holes \cite{Carr:1974nx,GarciaBellido:1996qt,Yoo:2020dkz,Kohri:2007qn}, which are expected to constitute at least a fraction of the cold dark matter \cite{Carr:2016drx,Bird:2016dcv,Sasaki:2016jop,Wang:2016ana,Authors:2019qbw,Carr:2020gox,Clesse:2016vqa,Eroshenko:2016hmn,Kashlinsky:2016sdv,Harada:2016mhb,Georg:2017mqk,Nakamura:2016hna,Sasaki:2018dmp,Vaskonen:2020lbd,DeLuca:2020agl}, and meanwhile explain the binary black holes observed by the advanced \ac{LIGO} \cite{LIGOScientific:2018mvr}. 

Gravitational waves have been proposed as a probe to the small-scale scalar perturbations \cite{Alabidi:2012ex,Alabidi:2013lya,Inomata:2018epa,Orlofsky:2016vbd,Ben-Dayan:2019gll,Kohri:2018awv,Wang:2019kaf,Kapadia:2020pir,Kapadia:2020pnr,Cai:2018dig,Yuan:2019wwo,Yuan:2020iwf,PhysRevD.81.023517,Bugaev:2010bb,Saito:2008jc,Saito:2009jt,Nakama:2016enz,Kohri:2020qqd,Jolicoeur:2018blf}. 
From a view of theoretical perspective, the gravitational waves could be non-linearly produced by the scalar perturbations in the second (or higher) order cosmological perturbation theory \cite{Mollerach:2003nq,Ananda:2006af,Baumann:2007zm,Assadullahi:2009jc}. 
Therefore, such a kind of gravitational waves is called the second order or induced gravitational waves, which are one of the important physical goals for the current or future gravitational-wave detection projects, for example, the \ac{LIGO}, Virgo and KAGRA detectors \cite{Aasi:2013wya}, the \ac{LISA} \cite{Audley:2017drz}, and the \ac{SKA} \cite{Moore:2014lga}. 

To measure the second order gravitational waves, one defines the energy density spectrum as the physical observable \cite{Allen:1997ad}. 
Upon the gauge fixing, such a spectrum has been widely investigated \cite{Inomata:2019yww,Yuan:2019fwv,DeLuca:2019ufz,Lu:2020diy,Tomikawa:2019tvi,Hwang:2017oxa,Giovannini:2020qta,Ali:2020sfw}.
Confusions arouse due to the contradictory claims across the literatures. 
There are two possible reasons to explain this issue. 
On the one side, it may be related to the gauge fixing, which could give rise to the fictitious tensor perturbations \cite{Giovannini:2020qta}. 
The induced gravitational waves have been shown to be gauge dependent \cite{Hwang:2017oxa,Tomikawa:2019tvi,Lu:2020diy}. 
To resolve this problem, one may represent them with the gauge invariant variables \cite{Bruni:1996im,Matarrese:1997ay,Malik:2008im,Nakamura:2006rk,Chang:2020tji,Domenech:2017ems,Gong:2019mui}. 
On the other side, it may be related to the physical observable of gravitational waves, which was suggested to be defined in  the synchronous frame \cite{DeLuca:2019ufz}. 
However, the authors argued that the gauge invariance is necessarily abandoned, since it is impossible to construct the gauge invariant second order synchronous variables \cite{Bertschinger:1993xt}. 

As will be shown, the gauge invariance could be reserved in studies of the second order gravitational waves. We can define well the gauge invariant variables for the tensor perturbations in the synchronous frame, though we ill-define the gauge invariant variables for the scalar and vector perturbations \cite{Chang:2020tji}. 
To be specific, in the gauge invariant synchronous frame, it can be shown that the gauge dependent second order gravitational waves are mixed with only the first order scalar perturbations that could be well defined in the gauge invariant Newtonian frame. 
In other words, the gauge invariant second order gravitational waves can be well defined in the synchronous frame, meanwhile not contaminated by the ill-defined gauge invariant synchronous scalar or vector perturbations. 
On the other side, it is obvious that the fictitious degrees of freedom are removed in the gauge invariant framework. 
Therefore, the gauge invariance should not be abandoned in the studies of the second order gravitational waves. 

In this \emph{Letter}, we will study the gauge invariance of the second order gravitational waves by following the Lie derivative method \cite{Matarrese:1997ay,Nakamura:2006rk,Chang:2020tji,Malik:2008im,Bruni:1996im}. 
We will present the gauge invariant synchronous variables and the equations of motion for the second order gravitational waves. In the equations of motion, the gravitational waves are sourced from the first order scalar perturbations that are defined in the gauge invariant Newtonian frame. 
Finally, the energy density spectrum of the second order gravitational waves will be defined in the gauge invariant synchronous frame, which is directly related to the measurements.

\emph{Gauge invariant perturbations of generic tensor to second order.}---Upon the first order gauge transformation {\ss $\tilde{x}=x+\zeta^{(1)}$}, the first order perturbation of a generic tensor {\ss $Q$} transforms as {\ss $\tilde{Q}^{(1)}=Q^{(1)}+\mathcal{L}_{\zeta_{1}}Q^{(0)}$} \cite{Bruni:1996im}, where {\ss $\mathcal{L}_{\zeta_{1}}$} is Lie derivative along an infinitesimal vector {\ss $\zeta_{1}=\zeta^{(1)}$} and {\ss $Q^{(0)}$} the tensor {\ss $Q$} on the background. 
It is straightforward to check that the quantity {\ss $Q^{(\mathrm{GI,1})}=Q^{(1)}-\mathcal{L}_{X}Q^{(0)}$} \cite{Nakamura:2019zbe}, where an infinitesimal vector {\ss $X$} transforms as {\ss $\tilde{X}=X+\zeta_{1}$}, are gauge invariant. 
Upon the second order gauge transformation {\ss $\tilde{x}=x+\zeta^{(1)}+\zeta^{(2)}/2$}, the second order perturbation transforms as
    {\ss $\tilde{Q}^{(2)}=Q^{(2)}+2\mathcal{L}_{\zeta_{1}}Q^{(1)}+(\mathcal{L}_{\zeta_{2}}+\mathcal{L}_{\zeta_{1}}^{2})Q^{(0)}$} \cite{Bruni:1996im}, 
where one defines an infinitesimal vector {\ss $\zeta_{2}=\zeta^{(2)}-\zeta^{(1)}\partial\zeta^{(1)}$}. 
It can be checked that the gauge invariant variable is {\ss $Q^{(\mathrm{GI},2)}=Q^{(2)}-2\mathcal{L}_{X}Q^{(1)}-(\mathcal{L}_{Y}-\mathcal{L}^{2}_{X})Q^{(0)}$} \cite{Nakamura:2019zbe}, where an infinitesimal vector {\ss $Y$} transforms as {\ss $\tilde{Y}=Y+\zeta_{2}+[\zeta_{1},X]$}. 
Due to infinite possibilities in the choice of {\ss $X$} and {\ss $Y$}, one could obtain infinite families of gauge invariant variables in principle.

\emph{Second order gravitational waves in terms of gauge invariant synchronous variables.}---Upon \ac{FLRW} metric {\ss $g^{(0)}_{\mu\nu}=a^{2}(\eta)\mathrm{diag}(-1,1,1,1)$}, the cosmological perturbations of $n$-th order are defined as {\ss $g_{00}^{(n)}=-2a^{2}\phi^{(n)}$}, {\ss $g_{0i}^{(n)}=g_{i0}^{(n)}=a^{2}(\partial_{i} b^{(n)} + \nu_{i}^{(n)})$} and {\ss $g_{ij}^{(n)} =a^{2}( - 2 \psi^{(n)} \delta_{ij} + 2\partial_{i} \partial_{j} e^{(n)} + \partial_{i} c_{j}^{(n)} + \partial_{j} c_{i}^{(n)} + h_{ij}^{(n)} )$}, 
where {\ss $\phi^{(n)}$}, {\ss $\psi^{(n)}$}, {\ss $b^{(n)}$} and {\ss $e^{(n)}$} are scalar, {\ss $\nu_i^{(n)}$} and {\ss $c_j^{(n)}$} vector, and {\ss $h_{ij}^{(n)}$} tensor. 
The decomposition is achieved by the transverse operator {\ss $\mathcal{T}^{i}_{j}=\delta^{i}_{j}-\partial^{i}\Delta^{-1}\partial_{j}$} \cite{Maggiore:2018sht}, where {\ss $\Delta^{-1}$} is the inverse Laplacian operator defined on the background. 
We have the transverse and traceless conditions, i.e., {\ss $\partial^{i} \nu^{(n)}_{i} = 0$}, {\ss $\partial^{i} c^{(n)}_{i} = 0$}, {\ss $\partial^{i} h^{(n)}_{ij} = 0$} and {\ss $\delta^{ij} h^{(n)}_{ij} = 0$}.
Here, {\ss $\eta$} and {\ss $a(\eta)$} denote the conformal time and the scale factor of the Universe, respectively.

It seems that there are ten independent modes in the cosmological perturbations of any order. 
However, four of them are unphysical due to the gauge transformations. 
We thus have only two independent modes, respectively, in the scalar, vector and tensor perturbations. 
Since the observable of gravitational waves is defined in the synchronous frame \cite{Maggiore:1999vm,DeLuca:2019ufz}, we expand the second order cosmological perturbations in terms of the gauge invariant synchronous variables, namely, {\ss $\Psi^{(2)}$}, {\ss $E^{(2)}$}, {\ss $C_{i}^{(2)}$} and {\ss $H_{ij}^{(2)}$}, 
for which we have {\ss $\partial^{i} C^{(2)}_{i} = 0$}, {\ss $\partial^{i} H^{(2)}_{ij} = 0$} and {\ss $\delta^{ij} H^{(2)}_{ij} = 0$}. 
It is known that there are residual gauge freedoms in the synchronous frame \cite{Bertschinger:1993xt}. 
If we represent the first order cosmological perturbations with the gauge invariant synchronous variables, the gauge freedoms arise at the source term in the equation of motion of {\ss $H_{ij}^{(2)}$} \cite{Lu:2020diy,Chang:2020tji}. 
However, they would not arise if we use the gauge invariant Newtonian frame at first order \cite{Chang:2020tji}. 
Meanwhile, the second order gravitational waves are decoupled from the second order scalar and vector perturbations. 
Therefore, we can define well the gauge invariant second order gravitational waves in the synchronous frame. 

Though we explore the second order gravitational waves in the gauge invariant synchronous frame, it is still allowed to study the first order cosmological perturbations in the gauge invariant Newtonian frame \cite{Chang:2020tji}. 
The gauge invariant first order Newtonian variables are defined by {\ss $\Phi^{(1)}=\phi^{(1)}-[a(e^{(1)\prime}-b^{(1)})]^{\prime}/a$}, {\ss $\Psi^{(1)}=\psi^{(1)}+\mathcal{H}(e^{(1)\prime}-b^{(1)})$}, {\ss $V^{(1)}_{i}=\nu^{(1)}_{i}-c^{\prime}_{i}$} and {\ss $H^{(1)}_{ij}=h^{(1)}_{ij}$} \cite{Mukhanov:1990me}, where {\ss $\mathcal{H}=a^{\prime}/a$} is the conformal Hubble parameter, and the prime denotes a derivative with respect to {\ss $\eta$}. 
These variables can be determined by {\ss $X^{0}=e^{(1)\prime}-b^{(1)}$} and {\ss $X^{i}=\delta^{ij}(c^{(1)}_{j}+\partial_{j}e^{(1)})$}. 
The scalar, vector and tensor perturbations are decoupled at first order, and therefore follow the equations of motion by themselves. 
This prediction is preserved for an arbitrary family of gauge invariant variables at first order, since the gauge transformation is governed by a vector rather than a tensor \cite{Riotto:2002yw}. 

Though the derivation of the second order gauge invariant variables is more complicated, we find that the gauge invariant second order synchronous tensor perturbations can be defined as  \cite{Chang:2020tji}
\be\label{eq:Hij}
    H_{ij}^{(2)}  =  h^{(2)}_{i   j} - \Lambda_{ij}^{kl} \mathcal{X}_{k   l}\ ,
\ee
where {\ss $\Lambda_{ij}^{kl}={\mathcal{T}^{l}_{i}\mathcal{T}^{m}_{j} - \mathcal{T}_{ij}\mathcal{T}^{lm}/2}$} denotes a transverse and traceless operator, and {\ss $\mathcal{X}_{\mu\nu}=1/a^{2}\mathcal{L}_{X}(2g^{(1)}_{\mu\nu}-\mathcal{L}_{X}g^{(0)}_{\mu\nu})$} (see explicit expressions in Appendix \ref{append}). 
The gauge invariant second order synchronous variables can be determined by 
{\ss $Y^{0}=1/a\int \mathrm{d}\eta(a\phi^{(2)}+a/2\mathcal{X}_{00})$} and {\ss $Y^{i}=\delta^{ij}\int\mathrm{d}\eta [\nu_{j}^{(2)}+\partial_{j}b^{(2)}+1/a\int\mathrm{d}\bar{\eta}(a\partial_{j}\phi^{(2)}) - \mathcal{X}_{0j}+1/(2a)\int\mathrm{d}\bar{\eta}(a\partial_{j}\mathcal{X}_{00})]$}. 
The second term at the right hand side of Eq.~(\ref{eq:Hij}) is uniquely determined by the first order cosmological perturbations. 
In other words, the gauge dependent second order gravitational waves are shown to mix with the first order cosmological perturbations, while decouple from the second order scalar and vector perturbations. 
We conclude that {\ss $H^{(2)}_{ij}$} is gauge invariant, while {\ss $h^{(2)}_{ij}$} is not. 
On the other side, if {\ss $H^{(2)}_{ij}$} is zero in a particular gauge, it will always be zero in any gauge. 
Therefore, we can immediately distinguish the physical second order gravitational waves from the fictitious ones. 
If {\ss $H^{(2)}_{ij}$} vanishes, the second order gravitational waves, if any, are fictitious and can be removed through the gauge transformations.

\emph{Equation of motion of the gauge invariant second order gravitational waves.}---The gauge invariant second order gravitational waves can be induced by the gauge invariant first order scalar perturbations. 
We could neglect the first order vector and tensor perturbations, i.e., {\ss $V^{(1)}_{i}=H^{(1)}_{ij}=0$}. If the perfect fluids were considered, we further have {\ss $\Phi^{(1)}=\Psi^{(1)}$} \cite{Mukhanov:1990me}. 
Adopting the Lie derivative method to the Einstein tensor and the energy-momentum tensor, we can obtain the $n$-th order gauge invariant Einstein tensor {\ss $G^{(n)}_{\mu\nu}$} and energy-momentum tensor {\ss $T^{(n)}_{\mu\nu}$} \cite{Nakamura:2006rk,Chang:2020tji,Nakamura:2019zbe}. 
The Einstein's field equation is {\ss $G^{(n)}_{\mu\nu}=\kappa~ T^{(n)}_{\mu\nu}$}, where {\ss $\kappa=8\pi G$} is constant and {\ss$G$} the gravitational constant. 
We could calculate {\ss $G^{(n)}_{\mu\nu}$} and {\ss $T^{(n)}_{\mu\nu}$} by replacing the metric with the gauge invariant one. 

The transverse and traceless part in the spatial component of the gauge invariant second order Einstein's field equations governs the equation of motion of the second order gravitational waves \cite{Chang:2020tji}
\be\label{eq:eom}
H^{(2)\prime\prime}_{\bold{k},ij}+2\mathcal{H}H^{(2)\prime}_{\bold{k},ij}+ {{k}}^{2} H^{(2)}_{\bold{k},ij} = 4 \Lambda^{lm}_{\bold{k},ij} {S}_{\bold{k},lm}\ ,
\ee
which has been expressed in the momentum space. 
Here, {\ss $H^{(2)}_{\bold{k},ij}$} is the Fourier mode of {\ss $H^{(2)}_{ij}$}, and the operator {\ss $\Lambda_{\bold{k},ij}^{lm}$} is the Fourier mode of {\ss $\Lambda_{ij}^{lm}$}, which is composed of {\ss $\mathcal{T}^{ij}_{\bold{k}}=\delta^{ij}-k^{i}k^{j}/k^{2}$}, i.e., the Fourier mode of {\ss $\mathcal{T}^{ij}$},. 
The source term {\ss ${S}_{\bold{k},lm}(\eta)$} is expressed in terms of the gauge invariant first order Newtonian scalar perturbations, namely,  \cite{Chang:2020tji}
\begin{widetext}
\be\label{eq:source}
{S}_{\bold{k},lm}(\eta) = \int \frac{d^{3}q}{(2\pi)^{3/2}} q_{l} q_{m} \Big(2\Phi^{(1)}_{\bold{q}}\Phi^{(1)}_{\bold{k-q}} + \frac{4}{3(1+w)} ({\mathcal{H}}^{-1}\Phi^{(1)\prime}_{\bold{q}}+\Phi^{(1)}_{\bold{q}}) ({\mathcal{H}}^{-1}\Phi^{(1)\prime}_{\bold{k-q}}+\Phi^{(1)}_{\bold{k-q}}) \Big)\ ,
\ee
\end{widetext}
where {\ss $w$} is the equation of state, and {\ss $\Phi^{(1)}_{\bold{q}}$} the Fourier component of {\ss $\Phi^{(1)}$}. 
In the derivation of the above equations, we use the zeroth and first order Einstein's field equations. 
One should note that all of the quantities are expressed in terms of the gauge invariant variables, in particular, {\ss $H^{(2)}_{ij}$} synchronous while {\ss $\Phi^{(1)}$} Newtonian. 
Therefore, there are not residual gauge freedoms in the equation of motion. 
Moreover, we take the equation of state to be {\ss $w=1/3$} (or {\ss $w=0$}) in the radiation (or matter) dominated epoch of the Universe. 

The equation of motion can be resolved with the Green's function method \cite{Mukhanov:1990me}. 
The solution is 
\be
H^{(2)}_{\bold{k},ij}=\frac{4\Lambda^{lm}_{\bold{k},ij}}{a(\eta)}\int^{\eta}\mathrm{d}\bar{\eta}\mathcal{G}_{\bold{k}}(\eta,\bar{\eta})a(\bar{\eta})S_{\bold{k},lm}(\bar{\eta})\ ,
\ee
where {\ss $\mathcal{G}_{\bold{k}}(\eta,\bar{\eta})$} is a solution of {\ss $\mathcal{G}^{\prime\prime}_{\bold{k}}+(k^{2}-a^{\prime\prime}/a)\mathcal{G}_{\bold{k}}=\delta(\eta-\bar{\eta})$}. 
We obtain {\ss $k\mathcal{G}_{\bold{k}}=\sin(x-\bar{x})$} in the radiation dominated epoch, while {\ss $k\mathcal{G}_{\bold{k}}=1/(x\bar{x})[(1+x\bar{x})\sin(x-\bar{x})-(x-\bar{x})\cos(x-\bar{x})]$} in the matter dominated one. 
Here, we denote {\ss $x=k\eta$} and {\ss $\bar{x}=k\bar{\eta}$}. 
Further, we desire to learn the evolution of {\ss $\Phi^{(1)}_{\bold{k}}$} which follows the master equation \cite{Mukhanov:1990me}
\be
\Phi^{(1)\prime\prime}_{\bold{k}}+\frac{6(1+w)}{(1+3w)\eta}\Phi^{(1)\prime}_{\bold{k}} + w k^{2} \Phi^{(1)}_{\bold{k}}=0\ .
\ee
For simplicity, we consider the adiabatic scalar perturbations. 
We define {\ss $\Phi^{(1)}_{\bold{k}}$} as a product of the transfer function {\ss $\Phi(x)$}, which approaches unity well before the horizon re-entry, and the primordial value {\ss $\Phi^{\mathrm{p}}_{\bold{k}}$}. 
Solving the master function, we obtain {\ss $\Phi(x)=9\sqrt{3}/x^{3}(\sin(x/\sqrt{3})-(x/\sqrt{3})\cos(x/\sqrt{3}))$} in the radiation dominated epoch, while {\ss $\Phi(x)=1$} in the matter dominated one. 
The power spectrum of the primordial value is defined by {\ss $\langle \Phi^{\mathrm{p}}_{\bold{k}} \Phi^{\mathrm{p}}_{\bar{\bold{k}}} \rangle=\delta(\bold{k}+\bar{\bold{k}})2\pi^{2}/k^{3}[3(1+w)/(5+3w)]^{2}\mathcal{P}_{\zeta}(k)$}, where {\ss $\zeta$} is the primordial curvature perturbation. 
We have already learnt a lot of information about {\ss $\mathcal{P}_{\zeta}$} on large scales \cite{Zyla:2020zbs}. 
By contrast, we know little about it on small scales \cite{Emami:2017fiy,Gow:2020bzo}. 
The induced gravitational waves are expected to bring new insights onto the small-scale phenomenology.

\emph{Energy density spectrum of second order gravitational waves.}---For the gravitational waves produced at {\ss $\eta$}, we define their energy density spectrum {\ss $\Omega_{\mathrm{GW}}$} \cite{Allen:1997ad} in an integral of {\ss $\rho_{\mathrm{GW}}(\eta) = \rho_{\mathrm{c}}(\eta) \int \Omega_{\mathrm{GW}}(k,\eta) \mathrm{d}\ln k$}, where {\ss $\rho_{\mathrm{GW}}(\eta)$} is the energy density of gravitational waves and {\ss $\rho_{\mathrm{c}}(\eta)=3\mathcal{H}^{2}/(\kappa a^{2})$} the critical energy density of the Universe at {\ss $\eta$}. 
Here, we follow the convention {\ss $H^{(2)}_{\bold{k},ij}=H^{(2)}_{\bold{k}\lambda}e^{\lambda}_{ij}$}, where {\ss $e^{\lambda}_{ij}$} denotes the polarization tensor, and {\ss $\lambda$} is a plus or cross mode \cite{Maggiore:1999vm}. 
For the second order gravitational waves, following the method in Ref.~\cite{Kohri:2018awv}, we have 
\be
\Omega_{\mathrm{GW}}(k,\eta) = \frac{k^{2}}{24{\mathcal{H}^{2}}} \overline{\mathcal{P}_{\mathrm{t}}(k,\eta)}\ ,
\ee
where the overline denotes an oscillation average \cite{Inomata:2016rbd}, and {\ss $\mathcal{P}_{\mathrm{t}}$} is the dimensionless power spectrum 
\be\label{eq:pt}
\langle H^{(2)}_{\bold{k}\lambda}(\eta) H^{(2)}_{\bar{\bold{k}}\bar{\lambda}}(\eta) \rangle = \delta_{\lambda \bar{\lambda}}\delta^{3}(\bold{k}+\bar{\bold{k}})\frac{2\pi^{2}}{k^{3}} \mathcal{P}_{\mathrm{t}}(k,\eta)\ .
\ee
In this way, we have defined the energy density spectrum with the gauge invariant synchronous variables.
Since the cosmos is expanding, the energy density spectrum today is given by \cite{Kohri:2018awv}
\be
\Omega_{\mathrm{GW},0}(k)=\frac{\Omega_{\mathrm{GW}}(k,\eta)}{\Omega_{\mathrm{r}}(\eta)}\Omega_{\mathrm{r}}(\eta_{0})\ ,
\ee
where {\ss $\Omega_{\mathrm{r}}=\rho_{\mathrm{r}}(\eta)/\rho_{\mathrm{c}}(\eta)$}, {\ss $\rho_{\mathrm{r}}(\eta)$} denotes the energy density of the relativistic matter, and {\ss $\eta_{0}$} the conformal time today. 
One should note {\ss $\Omega_{\mathrm{r}}(\eta)\simeq 1$} in the radiation dominated epoch of the Universe. 

\begin{figure}
    \includegraphics[width =\columnwidth]{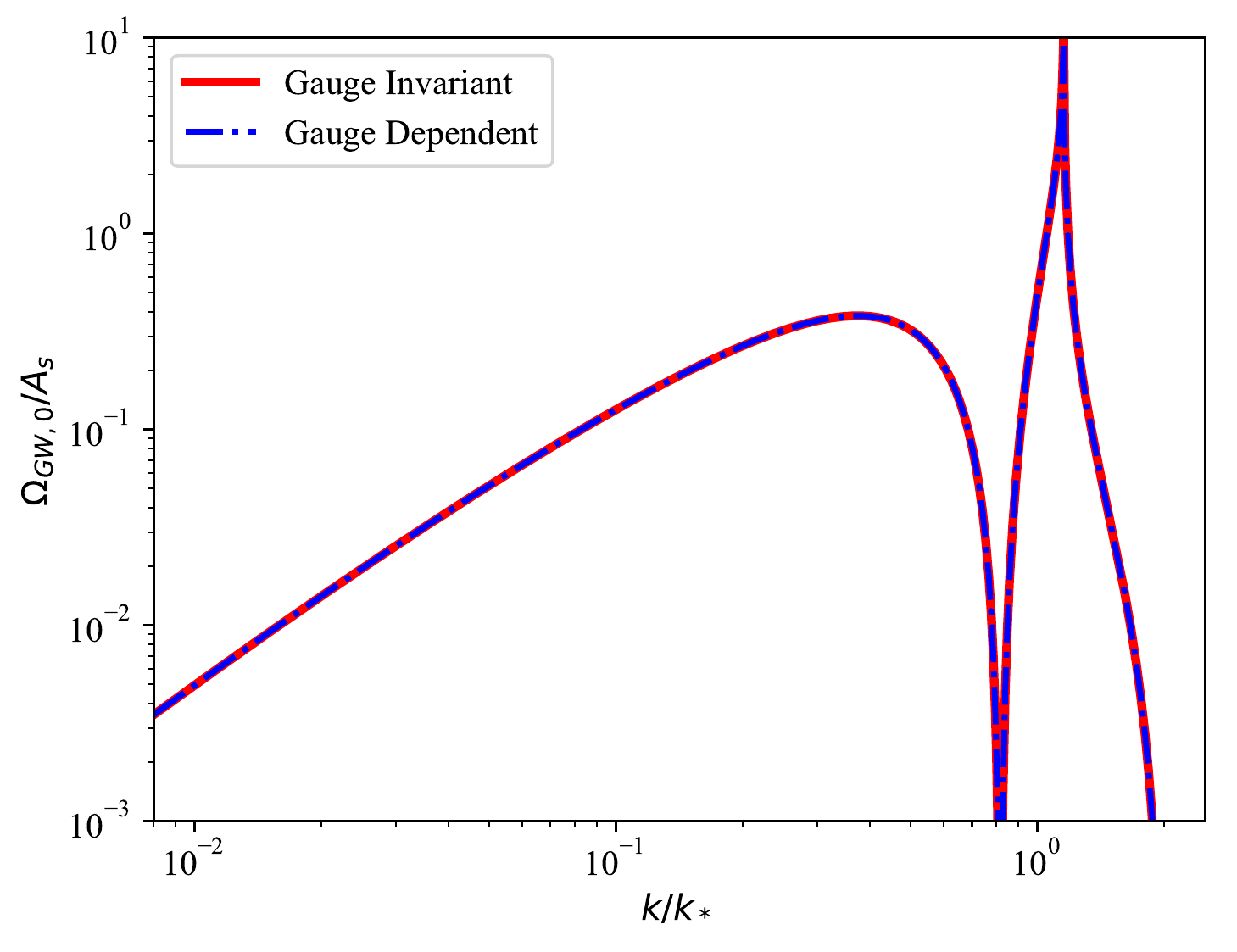}
    \caption{The present energy density spectra of the second order gravitational waves calculated in the gauge dependent frame and the gauge invariant frame, respectively.}\label{fig:ogwk}
\end{figure}

We could compare the gauge invariant {\ss $\Omega_{\mathrm{GW},0}$} with the previous gauge dependent one. 
Since the second order gravitational waves are induced by the first order scalar perturbations, we require the information about {\ss $\mathcal{P}_{\zeta}$}, in particular, on small scales. 
For illustration, the monochromatic curvature perturbations are considered as a typical scenario \cite{Kohri:2018awv,Wang:2019kaf}. 
The power spectrum is given by {\ss $\mathcal{P}_{\zeta}(k)=A_{s}\delta(\ln k - \ln k_{\ast})$}, where {\ss $A_{s}$} is normalization and {\ss $k_{\ast}$} a benchmark wavenumber at which the power spectrum is peaked. 
We focus on the second order gravitational waves produced in the radiation dominated epoch, since they have relatively high frequencies which are detectable by \ac{LIGO} \cite{Aasi:2013wya} and \ac{LISA} \cite{Audley:2017drz}. 
In Figure~\ref{fig:ogwk}, we show {\ss $\Omega_{\mathrm{GW},0}$} in the gauge invariant synchronous frame (red solid curve). 
We find that this spectrum coincides with that in the Newtonian gauge (blue dashed curve) \cite{Kohri:2018awv}. 
This result is not beyond our expectation.
In fact, Eqs.~(\ref{eq:eom})--(\ref{eq:pt}) would take the same form as those in the Newtonian gauge \cite{Kohri:2018awv,Inomata:2019yww}, if we simply replace the gauge invariant variables with the corresponding variables in the Newtonian gauge.  

In fact, although {\ss $H_{ij}^{(2)}$} is the gauge invariant synchronous variable, the gauge dependent components at the right hand side of Eq.~(\ref{eq:Hij}) could be calculated via choosing a gauge fixing. 
When the Newtonian gauge is fixed, based on the explicit expression of {\ss$\mathcal{X}_{kl}$} in Appendix, we find that {\ss$\mathcal{X}_{kl}$} vanishes since {\ss$b^{(1)}$}, {\ss$c_{i}^{(1)}$} and {\ss$e^{(1)}$} are zero. 
In other words, the second term at the right hand side of Eq.~(\ref{eq:Hij}) is zero. 
Therefore, we obtain {\ss$H_{ij}^{(2)}=h_{ij}^{(2,N)}$}, where {\ss$h_{ij}^{(2,N)}$} denotes the second order gravitational waves in the Newtonian gauge. 
This relation explains why one obtained correctly the energy density spectrum of second order gravitational waves from the Newtonian gauge.

\emph{Conclusion.}---In this \emph{Letter}, we represent the second order gravitational waves in the gauge invariant synchronous frame, where the physical observable is defined reasonably. 
The induced gravitational waves are found to be mixed with the lower order cosmological perturbations, but separated from the scalar and vector perturbations of the same order. 
In the gauge invariant framework, we derive the equation of motion of the second order gravitational waves, which are shown to be sourced from the first order scalar perturbations that are well defined in the gauge invariant Newtonian frame. 
Finally, we define the energy density spectrum of the second order gravitational waves in terms of the gauge invariant synchronous variable, implying no fictitious tensor perturbations remained. 
It is shown to coincide with the one in the Newtonian gauge.
The latter has been shown to be as same as the one in the synchronous gauge \cite{Inomata:2019yww,Yuan:2019fwv,Lu:2020diy,DeLuca:2019ufz}. 
Therefore, it is unnecessary to give up the gauge invariance in the investigations of the induced gravitational waves. 

One might wonder if the first order cosmological perturbations can be studied in other gauge invariant frames, rather than the Newtonian one. 
The answer is yes.
From the perspective of Lie derivative, we require a conversion formula to connect the gauge invariant synchronous-Newtonian variables considered in this work and another family of gauge invariant variables to be studied \cite{Chang:2020tji,Nakamura:2014kza}. 
We leave such a study for future works, since it is beyond the scope of this \emph{Letter}. 
On the other side, such a study may be related to the gauge dependence of the second order gravitational waves \cite{Lu:2020diy}. 
The second order gravitational waves have been shown to take the same energy density spectrum for some gauges, but divergent for others \cite{Inomata:2019yww,Yuan:2019fwv,DeLuca:2019ufz,Giovannini:2020qta,Tomikawa:2019tvi,Hwang:2017oxa,Lu:2020diy}. 
In fact, upon the gauge fixing, the energy density spectrum was defined in terms of the gauge dependent tensor perturbations {\ss$h_{ij}^{(2)}$} in the previous studies. 
However, as shown by us, it should be defined in terms of the gauge invariant synchronous variable {\ss$H_{ij}^{(2)}$}, i.e., a mixing between {\ss$h_{ij}^{(2)}$} and the first order scalar perturbations. 
For example, the energy density spectrum of {\ss$h_{ij}^{(2)}$} was shown to increase as {\ss$\eta^{6}$} in the uniform density gauge \cite{Lu:2020diy}. 
By contrast, as shown in Eq.~(\ref{eq:Hij}), there are counter terms in {\ss$H_{ij}^{(2)}$} to eliminate such an increase, and the energy density spectrum would be in coincidence with this work. 
We will explicitly study this point in a future work \cite{Chang:2020pre}, which is in preparation, since an explicit proof of it is also beyond the scope of this \emph{Letter}. 

Another concern is that the first order scalar perturbations are assumed to be adiabatic for simplicity. 
In fact, the method in this work is also available to other situations such as the isocurvature. 
We expect the derivation to be straightforward, but more complicated. 
In addition, the gauge invariant framework could be generalized to investigate the higher order gravitational waves in principle \cite{Nakamura:2014kza}. 

\vspace{0.3cm}
\begin{acknowledgements}
We acknowledge Prof. Tao Liu, Dr. Shi Pi, Mr. Chen Yuan and Mr. Jing-Zhi Zhou for helpful discussions. 
This work is supported by the National Natural Science Foundation of China upon Grant No. 12075249, No. 11675182 and No. 11690022, and by a grant upon Grant No. Y954040101 from the Institute of High Energy Physics, Chinese Academy of Sciences.
We acknowledge the \texttt{xPand} package \cite{Pitrou:2013hga}. 
\end{acknowledgements}

\bibliography{igw-gauge-invariance}

\begin{widetext}
\newpage

\appendix 
\section{Appendix: Expressions of $\mathcal{X}_{\mu\nu}$ and Gauge invariant second order variables}\label{append}
\noindent In this appendix, we briefly summarize the explicit expressions of $\mathcal{X}_{\mu\nu}$ and the gauge invariant second order cosmological perturbations \cite{Chang:2020tji}. \\

\noindent
{\ss \be
&&\mathcal{X}_{00}  =  2 \left\{ \frac{2}{a} \[a \(
  e^{(1)\prime} - b^{(1)}\)\]^\prime + \( e^{(1)\prime} - b^{(1)}\) \partial_0 +
  \delta^{ik} \(c_k^{(1)} + \partial_k e^{(1)}\) \partial_i \right\} \nonumber\\
  &  &\quad\quad\quad \times \left\{ 2
  \phi^{(1)} - \frac{1}{a} \[a \( e^{(1)\prime} - b^{(1)}\)\]^\prime \right\}
  - 2 \delta^{ik} \(c_k^{(1)} + \partial_k e^{(1)}\)^\prime  
  \(\partial_i b^{(1)} + 2 \nu_i^{(1)} - c_i^{(1)\prime}\)\ , \\
&&\mathcal{X}_{0 j} = - \left\{ \delta^k_j  \left[ 2\mathcal{H}
  \(e^{(1)\prime} - b^{(1)}\) + 
  \(e^{(1)\prime} - b^{(1)}\)^\prime+ \(e^{(1)\prime} - b^{(1)}\) \partial_0 +
  \delta^{il} \(c_l^{(1)} + \partial_l e^{(1)}\) \partial_i  \right] \right.\nonumber\\
  &  &\quad\quad\quad \left. + \delta^{ks} \partial_j \(c_s^{(1)} +
  \partial_s e^{(1)}\) \right\}
  \(\partial_k b^{(1)} + 2 \nu_k^{(1)} - c_k^{(1)\prime}\)  + 2
  \partial_j  \(e^{(1)\prime} - b^{(1)}\)  \left\{ 2 \phi^{(1)} -
  \frac{1}{a} \[a \(e^{(1)\prime} - b^{(1)}\)\]^\prime \right\}
  \nonumber\\
  &  &\quad\quad\quad - \delta^{kl} \(c_l^{(1)} + \partial_l e^{(1)}\)^\prime  \left\{ - 2
  \delta_{jk}  \left[ 2 \psi^{(1)} + \mathcal{H} \(e^{(1)\prime} -
  b^{(1)}\) \right] + \( 2 \partial_j \partial_k e^{(1)} + \partial_j c_k^{(1)} +
  \partial_k c_j^{(1)} \) + 2h_{jk}^{(1)} \right\}, \\
&&\mathcal{X}_{kl} = \left\{- \delta^s_k \delta^t_l  \left[ 2\mathcal{H}
  \(e^{(1)\prime} - b^{(1)}\) + \(e^{(1)\prime} - b^{(1)}\) \partial_0 +
  \delta^{ik} \(c_k^{(1)} + \partial_k e^{(1)}\) \partial_i \right] \right.  \nonumber\\
  &  &\quad\quad\quad \left. + \(\delta^s_k \delta^{tw} \partial_l + \delta^t_l \delta^{sw}
  \partial_k\)  \(c_w^{(1)} + \partial_w e^{(1)}\) \right\}
  \left\{ - 2 \delta_{st}  \left[ 2 \psi^{(1)} + \mathcal{H}
  \( e^{(1)\prime} - b^{(1)}\) \right]  \right. \nonumber\\
  &  &\quad\quad\quad \left. + \( 2 \partial_s \partial_t e^{(1)} +
  \partial_s c_t^{(1)} + \partial_t c_s^{(1)} \) + 2h_{st}^{(1)} \right\} 
  - \(\delta^s_k \partial_l + \delta^s_l \partial_k\)  \(e^{(1)\prime}
  - b^{(1)}\)  \(\partial_s b^{(1)} + 2 \nu_s^{(1)} - c_s^{(1)\prime}\)\ . \\\nonumber\\\nonumber\\
&&\Psi^{(2)} = \psi^{(2)} + \frac{\mathcal{H}}{a}  \int \mathrm{d} \eta ~a\( \phi^{(2)} + \frac{1}{2}
\mathcal{X}_{00} \)
  +\mathcal{T}^{kl} \mathcal{X}_{kl}\ , \\
&&{E}^{(2)} = e^{(2)} - \int \mathrm{d} \eta \[ b^{(2)} +
  \frac{1}{a}  \int \mathrm{d} \bar{\eta} \(a \phi^{(2)} \) \]  - \int \mathrm{d} \eta \[ \frac{1}{2 a}  \int \mathrm{d} \bar{\eta} \(a\mathcal{X}_{00} \)  - \Delta^{- 1} \partial^i \mathcal{X}_{0i} \]  \nonumber\\
&&\quad\quad\quad  - \frac{1}{4} \Delta^{- 1}  \(3 \Delta^{- 1} \partial^k \partial^l -
  \delta^{kl}\) \mathcal{X}_{kl}\ , \\
&&  {C}_i^{(2)} = c_i^{(2)} - \int \mathrm{d} \eta~ \nu_i^{(2)}
  +\mathcal{T}^j_i  \int \mathrm{d} \eta~ \mathcal{X}_{0 j} - \Delta^{- 1}
  \partial^j \mathcal{T}^k_i \mathcal{X}_{jk}\ , \\
&&  H_{ij}^{(2)} = h^{(2)}_{ij} - \( \mathcal{T}^k_i \mathcal{T}^l_j -
  \frac{1}{2} \mathcal{T}_{ij} \mathcal{T}^{kl} \) \mathcal{X}_{kl}\ . 
\ee}

\end{widetext}

\end{document}